# Research collaboration and the expanding science grid: Measuring globalization processes worldwide

Robert J.W. Tijssen[*,**,***], Ludo Waltman[*] and Nees Jan van Eck[*]

[*] *{tijssen, waltmanlr, ecknjpvan}@cwts.leidenuniv.nl*
Centre for Science and Technology Studies, Leiden University, Leiden (The Netherlands)

[**] Centre for Regional Knowledge Development, Leiden University, The Hague (The Netherlands)

[***] Centre for Research on Evaluation, Science and Technology, Stellenbosch University (South Africa)

**Abstract**
This paper applies a new model and analytical tool to measure and study contemporary globalization processes in collaborative science – a world in which scientists, scholars, technicians and engineers interact within a 'grid' of interconnected research sites and collaboration networks. The building blocks of our metrics are the cities where scientific research is conducted, as mentioned in author addresses on research publications. The unit of analysis is the geographical distance between those cities.

In our macro-level trend analysis, covering the years 2000–2010, we observe that research collaboration distances have been increasing, while the share of collaborative contacts with foreign cities has leveled off. Collaboration distances and growth rates differ significantly between countries and between fields of science. The application of a distance metrics to compare and track these processes opens avenues for further studies, both at the meso-level and at the micro-level, into how research collaboration patterns and trends are driving and shaping the connectivity fabric of world science.

**Introduction**
Science is an integral part of contemporary society that is engaged in a process of globalization, a rapidly changing environment in which individuals and institutions increasingly interact across territorial borders and at long distances. Academic science, engaged in basic research and conducted by footloose scholars, has always been at the forefront of this evolutionary development during the last two centuries. The production and utilization of scientific knowledge occurs through intermingled processes of creativity, communication, collaboration and competition. One of the connecting elements in this fabric of science is joint research between individuals and organizations at different locations (Royal Society, 2011). Some collaborate out of necessity, others because they expect added value from sharing resources and engaging in team work, or simply because it is a requirement set by research funding schemes. Small entities, with limited resources and facilities, are more inclined to engage in cross-border collaboration than those who can rely on an adequate local science base or national infrastructure. The geographical distance between research partners is one of the defining background characteristics of these collaborative relationships and networks, which may arise from informal small-scale collaborative linkages between close colleagues to formalized and resource-intensive large-scale international collaborations on large facilities such as the International Space Station, the Large Hadron Collider at CERN, or the international fusion reactor project (ITER). With the alleged

'death of distance' (Cairncross, 1997), physical proximity would appear to become less of a defining factor, with ever more possibilities to collaborate on a worldwide scale nowadays at relatively low cost. Even so, empirical studies continue to show that the propensity to seek collaborative partners close by, either locally or domestically, remains a major determining factor in academic science (e.g., Hennemann et al., 2011; Hoekman, 2010; Jones et al., 2008; Wuchty et al., 2007).

Collaboration patterns and trends emerge from a highly complex adaptive system with millions of largely unseen interactions and transactions, large uncertainties about when and how people work together, significant effects of chance meetings (serendipity) and major influences of external factors such as funding mechanisms, managerial decisions or policy interventions. Numerous interconnected determinants are at play of which only a few major structural factors can be modeled at the macro-level of national research systems: Geographical borders and language, fields of science, research capacity and human resources, stage of economic development, international orientation and the existence of major research-performing universities, government laboratories or R&D-intensive companies.

Macro-level studies of science indicate that the average distance between co-authoring research partners has indeed gradually and continuously increased over the last decades (Tijssen et al., 2011; Waltman et al., 2011). In that sense, one could argue that science is in a process of globalization in terms of trans-national research collaboration agreements, shared resources, joint activities, migration of people and flows and exchanges of knowledge and skills. These shifts and integration processes occur within a world-spanning evolutionary framework that encompasses a myriad of informational, infrastructural, political, cultural, social and economic factors affecting structures, patterns and trends in collaboration in science.

One of the key determinants of globalization in science is the availability of potential partners and willingness to collaborate on joint research activities. With the opening of new regions and countries into the world economy, and the influx of new entrants in worldwide science, especially from economically emerging nations, the likelihood to finding novel long-distance possibilities for collaboration has increased significantly in recent years. Striking a parallel with large-scale interconnected technical infrastructures, like electric power grids, one could say that this imaginary structure of possible collaborative research linkages, this 'grid of science', is expanding.

Defining geographical entities as the physical nodes on this grid of science, we can introduce a distance metrics between these entities to systematically study structural and temporal properties of collaborative linkages. A distance-based analytical model covers a spectrum of entities: From within-city 'local' collaboration to long-distance 'global' collaboration, with 'regional' and 'domestic' as intermediate categories. Using cities as a unit of analysis also introduces the option of aggregation, notably to intra-country regions, countries or even entire continents. These aggregate units can be monitored and compared in terms of their interconnectedness and degree of globalization.

This study aims at producing a view of broad patterns and general trends worldwide. We realize that these trends and patterns are significantly determined by country-specific determinants, either in terms of countries' science systems or external factors such as geo-political or economic developments. Hence, our analysis incorporates both the entire world as an aggregate and a limited number of countries that are seen as major contributors to world science: Australia, Canada, France, Germany, Japan, Netherlands, South Korea, United Kingdom and US, as well as the BRICS group of countries (i.e., Brazil, Russia, India, China and South Africa). Building on our previous work (Waltman et al., 2011), we also examine globalization trends at the level of broad fields of science.

The following general research questions will be addressed in this paper:
1. How does the actual growth rate of collaboration distances compare to the growth rate in a hypothetical fully globalized world?
2. Do the general trends in domestic collaboration distances differ from those in international collaboration distances?
3. To what extent is globalization in science a field- or country-dependent phenomenon?

The remainder of this paper is structured as follows. We first introduce our data source and methodology. We then present our empirical findings and their interpretation with regard to the research questions. Finally, we comment on issues for further research.

**Data source and methodology**
Research collaboration patterns and trends are documented by the hundreds of thousands of research publications that are published in the open scientific literature each year. The vast majority of these publications contain author affiliate addresses, where collaboration leaves a trail behind through multiple author addresses within jointly authored collaborative works. Our measurement model is based on the villages, towns and cities that are mentioned in these addresses, which collectively will be referred to as 'cities' from now on. Adopting this particular geographical perspective and level of aggregation, a city can be seen as a producer of research publications, a distinct entity in the landscape of science.

Our data source is Thomson Reuters' Web of Science (WoS) database. We took into account all 11.1 million WoS-indexed publications that appeared between 2000 and 2010, that are of the document type 'article' or 'review' and that have at least one address in their address list. For each publication, we deduplicated addresses that occur multiple times in the address list. It is important to note that we use the term 'address' to refer to a combination of a city and a country (and in the case of the US and Canada also a state) and that we ignore other address elements, such as an organization name, a street or a postal code. Overall, we obtained 18.8 million addresses. For each of these addresses, we tried to identify the corresponding geographical coordinates (i.e., latitude and longitude). The geocoding approach that we followed is identical to the approach that we took in our earlier work (Waltman et al., 2011). Using this approach, we were able to identify the geographical coordinates of 98.3% of all addresses. Addresses for which the geographical coordinates could not be identified were disregarded in the calculations of geographical distances discussed below.

Based on the above described data, calculations were performed both for a number of selected countries and for the world as a whole. The following five measures were calculated:
- *Proportion collaborative publications*. The proportion of the publications in which a country participates that are of a collaborative nature. A publication is considered collaborative if it has at least two addresses.

- *Proportion international collaborative relations.* The proportion of the collaborative relations in which a country participates that are international.
- *Mean geographical collaboration distance of all collaborative relations (MGCD All).* The average geographical distance between collaborating cities calculated over all collaborative relations in which a country participates.
- *Mean geographical collaboration distance of domestic collaborative relations only (MGCD Domestic).* The average geographical distance between collaborating cities calculated over all domestic collaborative relations in which a country participates. Hence, the two cities must be in the same country.
- *Mean geographical collaboration distance of international collaborative relations only (MGCD International).* The average geographical distance between collaborating cities calculated over all international collaborative relations in which a country participates. Hence, the two cities must be in different countries.

Note that MGCD All equals the proportion of domestic collaborative relations times MGCD Domestic plus the proportion of international collaborative relations times MGCD International. Hence, MGCD All can be seen as a weighted average of MGCD Domestic and MGCD International, where the weighting is determined by the proportion domestic and international collaborative relations of a country.

A fractional counting approach was applied in the calculation of the above measures, both at the level of publications and at the level of collaborative relations. The following example illustrates this. Suppose we have three publications, X, Y and Z. Publication X has two addresses from country A, one from country B and one from country C; publication Y has three addresses from country A and no other addresses; publication Z has one address from country A and no other addresses. In our fractional counting approach, publication X is assigned to country A with a weight of 1/2 and to countries B and C with a weight of 1/4 each. Publication X involves 4 * (4 − 1) / 2 = 6 collaborative relations. Each of these relations has a weight of 1/6. There is one domestic collaborative relation (between the two addresses in country A), and there are five international collaborative relations. Country A participates in four of the five international collaborative relations. Publication Y is fully assigned to country A. This publication involves 3 * (3 − 1) / 2 = 3 collaborative relations, all of them domestic. Each of these relations has a weight of 1/3. Like publication Y, publication Z is fully assigned to country A, but unlike publication Y, publication Z does not involve any collaboration. If publications X, Y and Z are the only publications from country A, then country A's proportion collaborative publications equals (1/2 + 1) / (1/2 + 1 + 1) = 3/5. Country A's proportion international collaborative relations then equals (4 * 1/6) / (5 * 1/6 + 3 * 1/3) = 4/11.

To investigate to what degree science is globalized, the calculations discussed above can also be performed for a hypothetical fully globalized world. The real world of science can then be compared with this hypothetical world. In this paper, we consider a hypothetical world that is based on a simple random collaboration model. Two core assumptions are made in this model: (1) Researchers everywhere in the world are equally likely to engage in collaboration, and (2) collaboration partners choose each other randomly, without paying attention to the geographical distance between them.[1] Our random collaboration model can be illustrated as follows. Suppose that in the real world we have three addresses, A, B and C, and that these addresses have produced, respectively, 10, 20 and 60 publications. Based on the assumptions of our random collaboration model, one would expect there to be three times as many collaborative relations between addresses A and C than between addresses A and B. Similarly, one would expect there to be six times as many collaborative relations between addresses B and C than between addresses A and B. What this means is that of all collaborative relations in our random collaboration model 10% is expected to be between A and B, 30% is expected to be between A and C, and 60% is expected to be between B and C. The calculations discussed above for the real world can also be performed for the hypothetical fully globalized world based on our random collaboration model. Comparing the outcomes obtained for the real world with those obtained for the hypothetical world yields an indication of the degree to which the real world is globalized.

---

[1] Our random collaboration model is based on similar ideas as the geographical dispersion measure that we used in our earlier research (Waltman et al., 2011). Similar ideas are also used in the work of Hennemann et al. (2012).

In the discussion of our empirical findings in the next section, we use the term 'random geographical collaboration distance' (RGCD) to refer to the mean geographical collaboration distance calculated based on our random collaboration model.

**Empirical findings**

*General trends*
Research collaboration is becoming increasingly common in world science. The share of collaborative publications progressed from 35.0% to 45.6% between 2000 and 2010. However, the share of international collaborative relations, involving cities in two different countries, has remained fairly stable during those years, fluctuating between 40.8% and 41.8%. In other words, research collaboration across cities is still increasing, but the share of international collaborations has leveled off. Based on this observation alone, one could jump to the conclusion that globalization has come to a halt. But this would ignore the fact that geographical distances between co-publishing research partners have gone up significantly, as shown in Figure 1. MGCD All has risen by 5.2% during the years 2000–2010 from 2 806 to 2 951 kilometers.

If cities were to distribute their collaborative relations randomly across partner cities that are active in scientific research, MGCD All would have been equal to RGCD All, which stretched to 7 171 kilometers in 2010. RGCD All went up by 7.7% in the last decade, thus outpacing actual collaboration distances. Clearly, we are in a stage of expansion, a process where cities are either increasingly engaged in long-distance collaborations with existing partners (intensifying established long-distance collaborations), or opening up collaborative links with new cities on the global science grid that are located at a longer distance than the previous average collaboration distance (new long-distance collaborations).

In Figure 1, both MGDC and RGCD have been split according to the type of collaboration – either domestic or international. Focusing on international partners, RGCD International rose to 7 696 kilometers between 2000 and 2010 with a growth of 4.8%. MGCD International grew slightly more, by 6.6%, to 6 126 kilometers, a clue that actual globalization processes are benefitting from an expanding science grid, from new opportunities for partnering with cities at longer distances. In contrast, at the domestic level, MDCG Domestic and RGCD Domestic both show a decline of about 2%, which suggests a tendency within countries to engage in collaborations with partner cities closer by.

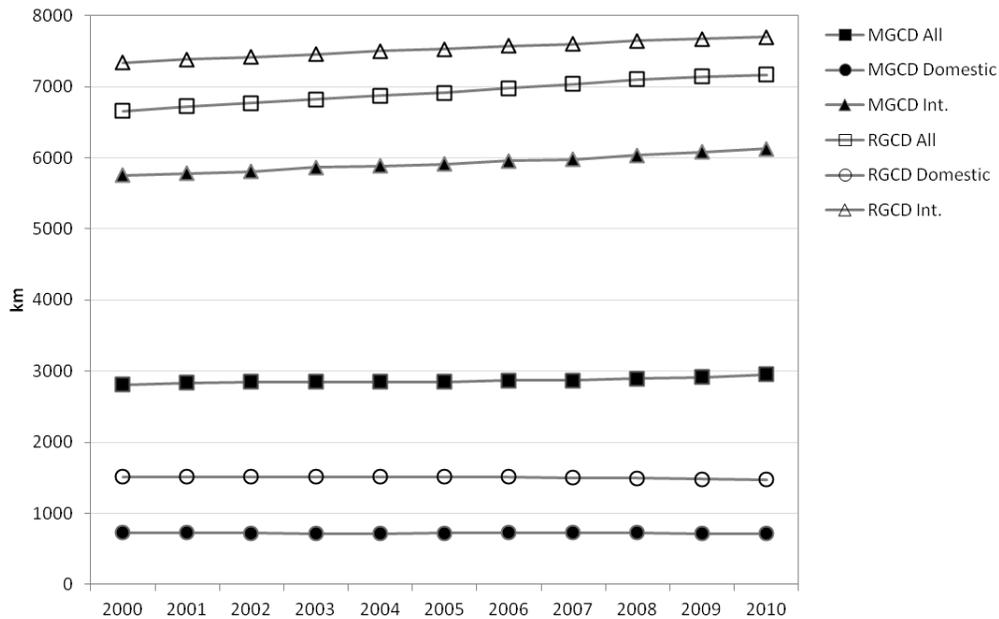

Figure 1. Trends in actual collaboration distances (MGCD) and random collaboration distances (RGCD).

In an age of globally interconnected science where collaboration, shared research methods and digital communication are increasingly commonplace, one might have expected to see a convergence of MGDC and RGCD. Figure 1 shows us otherwise. In 2010, RGCD All was 2.43 times larger than MGCD All. Hence, current collaboration patterns far from exhaust the collaboration possibilities according to our random collaboration model. Domestic collaboration patterns reveal a similar gap between MGCD and RGCD. RGCD Domestic is 2.06 times larger than MGCD Domestic. In the case of international collaboration, we find MGCD and RGCD to be more comparable, indicating that cross-border collaboration is relatively insensitive to geographical distance. The trends in Figure 1 also indicate that gaps between MGCD and RGCD were persistent throughout the last decade.

*Trends by field of science*
The general trends and patterns depicted in Figure 1 hide well-known differences among fields of science. Each field is characterized by its own historically determined set of knowledge production parameters, communication practices and governance systems that collectively determine the shape, size and dynamics of research collaboration patterns. For example, owing to the use of shared scientific instruments, astronomers by necessity display a greater propensity for long-distance collaboration than say civil engineers or sociologists, as shown in our previous study (Waltman et al., 2011).

To what degree are fields of science a critical parameter in our measures of globalization? Table 1 displays summary statistics for world science (ALL) as well as for four broad fields of science: Natural sciences (NAT), life, medical and health sciences (LIFE), Engineering science, computer science and mathematics (ENG), and social and behavioral sciences (SOC).[2] Collectively these four fields cover the entire WoS, with the exception of the arts and humanities and multidisciplinary journals such as *Nature* and *Science*. With an MGCD All of 3 236 kilometers, ENG on average involves collaboration over the longest distances. LIFE has the shortest average collaboration distance, with a 14% lower MGCD All than ENG. Notice however that the proportion collaborative publications is considerably higher in LIFE than in ENG (48% vs. 43%). Furthermore, LIFE leads in terms of RGCD All, indicating that compared with the other broad fields LIFE research is most geographically dispersed.

As for the general trends by field, all fields show an increase in both MGCD All and RGCD All. SOC is by far increasing fastest: MGCD All went up by 16% and RGCD All by 17%. Interestingly, MGCD International declined by 2% and MGGD Domestic by 8%. Hence, the growth in MGCD All is completely due to the large increase (by 32%) in the proportion international collaborative relations. NAT and LIFE both show significant overall growth rates. ENG is slightly lagging with just a few percent growth between 2000 and 2010, which might indicate a leveling-off effect. As for the share of collaborative publications and the proportion international collaborative relations, we find NAT lagging behind the other fields in terms of low growth rates, which again might refer to emerging leveling-off processes.

Table 1. Trends in actual collaboration distances (MGCD) and random collaboration distances (RGCD): Distances in 2010 (in kilometers) and relative change between 2000 and 2010 (within parentheses, in %).

|  | ALL | ENG | LIFE | NAT | SOC |
|---|---|---|---|---|---|
| MGCD All | 2951 (5) | 3236 (2) | 2768 (6) | 3032 (5) | 3034 (16) |
| RGCD All | 7171 (8) | 7008 (3) | 7168 (10) | 6981 (4) | 6546 (17) |
| MGCD Domestic | 715 (-2) | 680 (-8) | 699 (0) | 680 (0) | 933 (-8) |
| RGCD Domestic | 1474 (-2) | 1304 (-13) | 1528 (1) | 1271 (-3) | 1588 (-3) |
| MGCD Int. | 6126 (7) | 6264 (2) | 6224 (6) | 5861 (10) | 6485 (-2) |
| RGCD Int. | 7696 (5) | 7421 (1) | 7774 (6) | 7433 (4) | 7604 (3) |
| % Collab. publications | 46% (30) | 43% (25) | 48% (38) | 47% (20) | 41% (39) |
| % Int. collab. relations | 41% (0) | 46% (2) | 37% (2) | 45% (-4) | 38% (32) |

*Trends by country*
Collaboration patterns and processes are likely to be country-specific up to a certain degree. Small and less-developed countries may be more likely to engage in cross-border collaboration than large advanced countries. Geographically isolated countries need to collaborate across longer distances than countries that are well positioned within science-intensive continents or territories. A fuller understanding of globalization processes worldwide requires insight into the collaboration patterns and trends of individual countries. Our comparative analysis covers a selection of 14 countries, all of which are OECD member or have a partnership with the OECD. This set represents several of the world's leading economies and many of the largest nations in terms of science.

---

[2] These four broad fields were created by grouping together WoS subject categories based on citation patterns. This was done using the VOSviewer software (www.vosviewer.com). The final allocation of subject categories to the four broad fields was done semi-automatically. Background information on the scope of the different categories was taken into account as well.

Figure 2 depicts the proportion collaborative publications and the proportion international collaborative relations for each of the 14 countries, where the world average score is added as a general reference. There is a clear divide into two distinct subsets according to a country's share of international collaborative relations, that is, collaborative relations involving a research partner in a city abroad: 20–35% in the 'domestically oriented' countries and 45–60% in the 'internationally oriented' countries. Figure 3 adds the corresponding trend data. Several countries exhibit a diminishing share of international collaborative relations. Most of these declines occur within the group of domestically oriented countries. Brazil is an extreme case, with a very significant drop in the share of international collaborative relations.

Figure 4 presents each country's MGCD Domestic and MGCD International in 2010. The differences among countries in terms the geographical distances are significant. In the case of the Netherlands, domestic partners are on average within 100 kilometers from each other, while international partners are at 3 500 kilometers on average. At the other extreme, Australian researchers face domestic and international collaboration distances of 1 000 and 13 000 kilometers, respectively. The science systems in Canada, China, Russia and the US are also characterized by long distances between domestic partners, with averages of more than 800 kilometers.

Figure 5 displays the corresponding growth rates over the years 2000–2010. It divides the countries into the following four quadrants:
- Domestic and international expansion of collaboration distances (Canada and France are the most extreme cases in this quadrant).
- Domestic contraction and international expansion (China).
- Domestic expansion and international contraction (Australia, Japan and South Africa).
- Domestic and international contraction (India).

The increasing domestic and international collaboration distances of France indicate that research partnering occurs at increasingly long distances. Although the increase in MGCD Domestic is just 20 kilometers, it might be a signal that France is slightly moving away from its traditional Paris-centered science system. In the case of China, MGCD International increased substantially, but MGCD Domestic went down. This trend reflects a simultaneous process of enhancing domestic collaboration with partners at relatively close proximity, while engaging in foreign partnerships at longer distances (Liang & Zhu, 2002).

Domestic partner distances are on the rise in Australia, Japan and South Africa, while distances to foreign partners are diminishing. In the case of Australia and South Africa, this trend might reflect a diminishing dependence on partners in the northern hemisphere.[3] India shows double contraction, where collaboration distances went down both domestically and internationally. In a globalizing world of science, this deviant pattern could easily be interpreted as a structural weakness within the Indian science system – the inability among Indian researchers to keep up with other countries in successfully engaging in long-distance collaborations. However, a more detailed analysis of our data reveals that this outcome is at least partly due to changes in the WoS database. It turns out that in the period 2004–2010 about 30 local Indian journals, defined by us simply as journals with 'Indian' in their title, were added to the database. Most likely, the influx of publications from these journals deflates the distances between research partners, because these publications will mostly contain author addresses in India.

---

[3] South Africa differs from most other African nations which still rely heavily on north-south research partnerships, and associated influx of resources, to sustain or improve their underdeveloped science systems (Nordling, 2012).

**Further research is needed**
Research collaboration is without doubt a pervasive process in contemporary science. Our empirical findings shows a continuing increase in the geographical distances between collaborating partners. Basically, it could indicate either that long-distance partners participate in a larger share of all collaborative relations or that researchers are establishing new links with partners located at longer distances than previously. In the macro-level analysis reported in this paper, we did not attempt to disentangle these two processes. This would require a more detailed analysis at the meso-level of individual cities or at the micro-level of individual research institutions.

Our analysis also revealed marked differences among countries. Some countries are clearly contributing to scientific globalization, while others are not. The latter group may include countries with significant numbers of local journals in the WoS database (either journals in the local language or English language journals targeting a local readership). The presence of such journals in our database may have affected our findings, for instance in the case of India. Further studies should take note of possible database effects, and if necessary correct for them.

Overall, our findings provide ample evidence that globalization processes in science are on-going and that both country-specific and field-specific mechanisms are at play. The interaction between these two structural factors is an interesting topic for further research. At this point, we can only speculate about the most likely generic determinants underlying our findings. We would like to suggest three candidates for further empirical research:

- *Proximity between research partners*. Researchers are, like most humans, prone to proximity effects. They prefer to seek their partners close by, which correlates with shared language, culture and research funding systems, all of which increase the likelihood of successful collaboration resulting in jointly authored research publications (Frenken et al., 2010).
- *Policies and funds to support international collaboration and participation in global research networks*. The presence of such framework conditions may boost or bust the motivation among scholars, scientists, engineers and technicians to engage in long-distance collaborations. More detailed studies, especially at the level of countries or regions, should try to take into account the effects of national or supranational policy initiatives on non-local collaboration patterns and trends. For example, in the case of Europe, the European Commission's upcoming *Horizon 2020* program is likely to be major contributing factor to the expansion of intra-European collaboration, with spillover effects as to where European researchers seek partners outside their continent.
- *Concentration of science and technology resources among cities worldwide*. Some research sites are much larger than others (Florida, 2005; Leydesdorff & Persson, 2010). In our database, we find that fewer than 200 cities worldwide, each with major universities and institutes, are sufficient to account for more than half of all scientific output. Owing to decades or even centuries of scientific development and the accumulation of physical and human capital, such core nodes in global science are likely to attract a relatively large number of partners for collaboration.

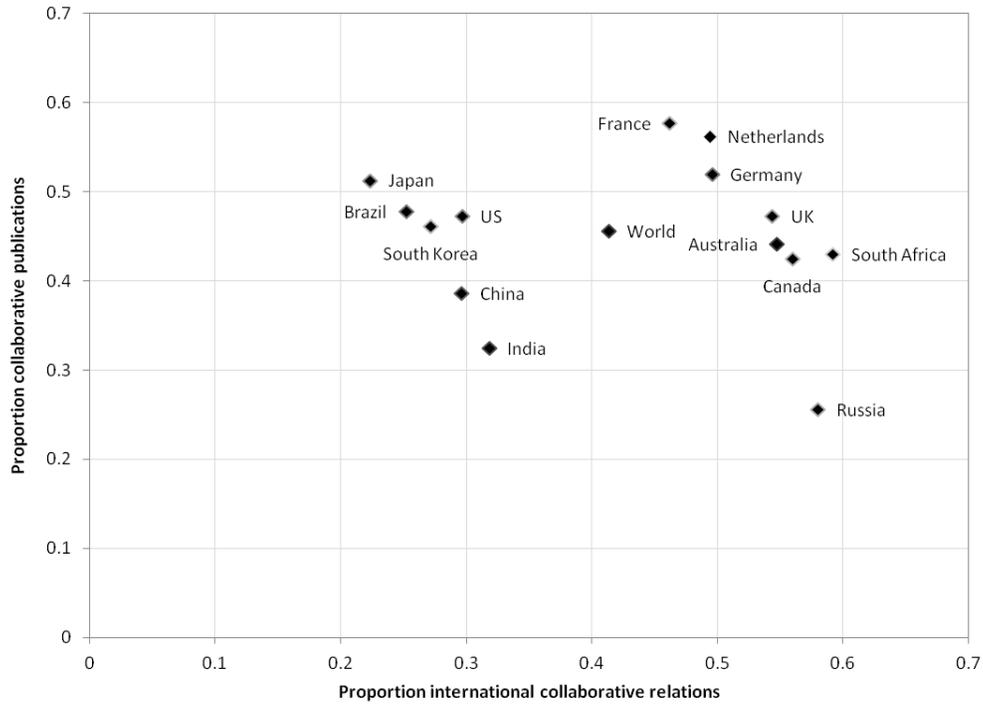

Figure 2. Proportion collaborative publications and proportion international collaborative relations for 14 selected countries, based on publications in 2010.

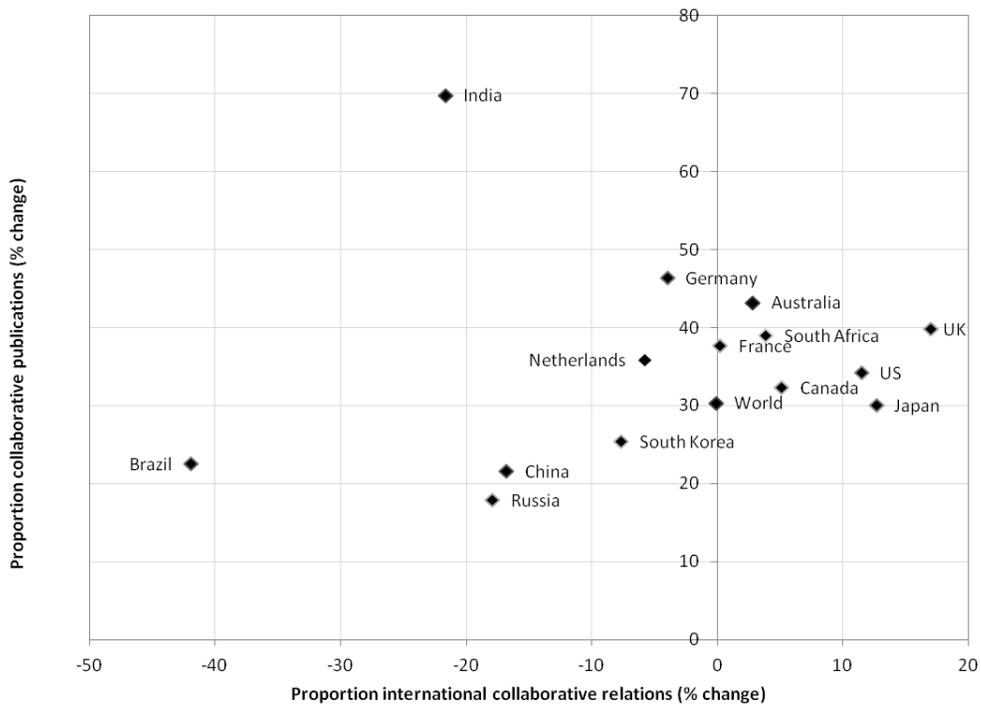

Figure 3. Relative change between 2000 and 2010 in the proportion collaborative publications and the proportion international collaborative relations for 14 selected countries.

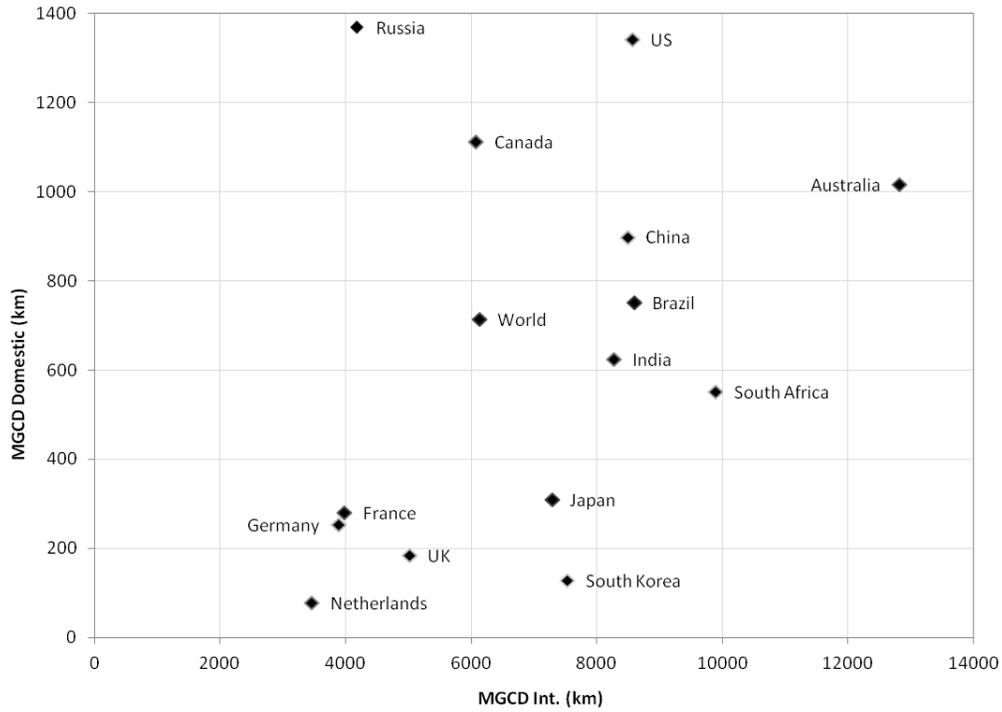

Figure 4. MGCD Domestic and MGCD International for 14 selected countries, based on publications in 2010.

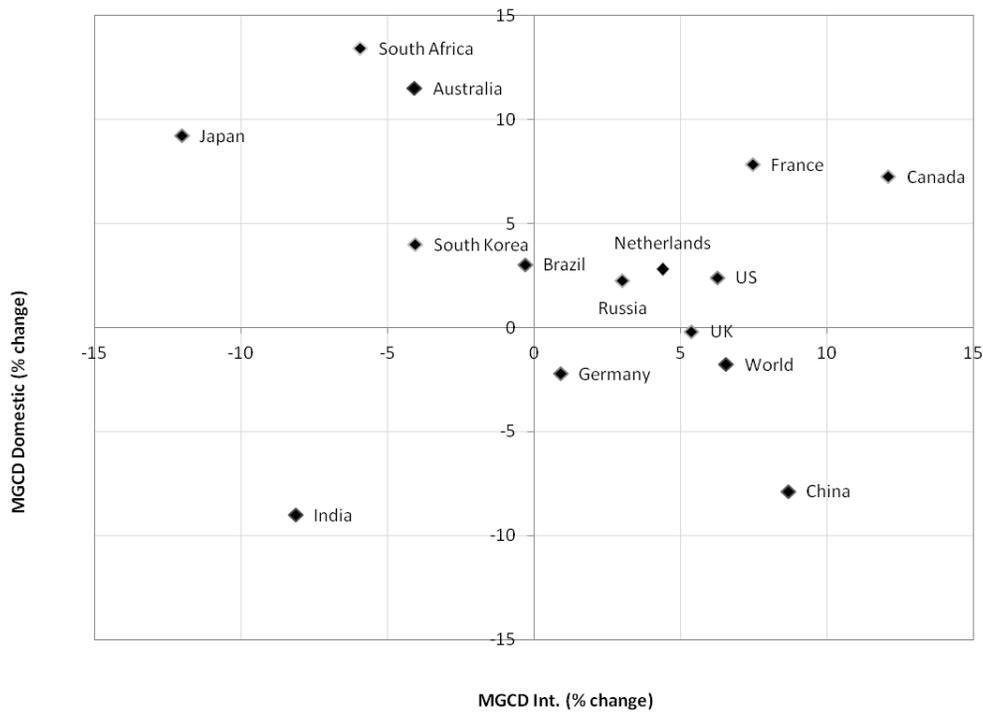

Figure 5. Relative change between 2000 and 2010 in MGCD Domestic and MGCD International for 14 selected countries.